\documentclass[superscriptaddress,nofootinbib,twocolumn,prb]{revtex4-1}

\usepackage{amsmath,graphicx}
\usepackage{color}

\newcommand{\rr}{{\bf r}}

\newcommand{\dr}{{\textrm d}{\bf r}}

\begin{document}

\title{The standard mean-field treatment of inter-particle attraction in classical DFT is better than one might expect}
\author{Andrew J. Archer}
\affiliation{Department of Mathematical Sciences, Loughborough University, Loughborough, LE11 3TU, UK}
\author{Blesson Chacko}
\affiliation{Department of Mathematical Sciences, Loughborough University, Loughborough, LE11 3TU, UK}
\author{Robert Evans}
\affiliation{H. H. Wills Physics Laboratory, University of Bristol, Bristol, BS8 1TL, UK}

\begin{abstract}
In classical density functional theory (DFT) the part of the Helmholtz free energy functional arising from attractive inter-particle interactions is often treated in a mean-field or van der Waals approximation. On the face of it, this is a somewhat crude treatment as the resulting functional generates the simple random phase approximation (RPA) for the bulk fluid pair direct correlation function. We explain why using standard mean-field DFT to describe inhomogeneous fluid structure and thermodynamics is more accurate than one might expect based on this observation. By considering the pair correlation function $g(x)$ and structure factor $S(k)$ of a one-dimensional model fluid, for which exact results are available, we show that the mean-field DFT, employed within the test-particle procedure, yields results much superior to those from the RPA closure of the bulk Ornstein-Zernike equation. We argue that one should not judge the quality of a DFT based solely on the approximation it generates for the bulk pair direct correlation function.
\end{abstract}

\maketitle

\section{Introduction}

Classical density functional theory (DFT)\cite{evans1979nature, evans1992density, lowen2002density, lutsko, hansen2013theory} is based on the idea that the thermodynamic grand potential of an inhomogeneous fluid can be expressed as a functional of the average one-body density profile $\rho(\rr)$. Minimizing an (approximate) functional with respect to $\rho(\rr)$ provides an estimate of the equilibrium density profile and the grand potential. DFT has proved to be a versatile tool for determining the thermodynamic quantities arising in the physics of adsorption and fluid interfaces. These include, for example, interfacial tensions\cite{evans1992density} and the solvation force (excess pressure) for confined fluids.\cite{evans1987phase, stewart2014layering} Since DFT provides directly the grand potential it is particularly well-suited to investigations of surface phase behaviour and perhaps it is here where DFT has had most success, revealing new phenomena and new physical insight.\cite{evans1992density, hansen2013theory} DFT also provides a direct measure of density fluctuations in the inhomogeneous fluid via the local compressibility $(\partial \rho(\rr)/\partial\mu)$, where $\mu$ is the chemical potential and recent papers have investigated this quantity for liquids at solvophobic planar substrates\cite{evans2015local} and confined between a variety of nanostructures.\cite{chacko2017solvent} In addition to the one-body density, higher order correlation functions can be obtained by taking further functional derivatives of the free energy functional. In particular, the two-body pair direct correlation function $c^{(2)}(\rr,\rr')$ is proportional to the second functional derivative of the excess Helmholtz free energy with respect to the density.\cite{evans1979nature, evans1992density, lutsko, hansen2013theory} It is tempting to assess the quality of an approximate DFT for a given model fluid by comparing the resulting $c^{(2)}(r)$ for a homogeneous fluid with that obtained from computer simulation {of the radial distribution function $g(r)$ or structure factor $S(k)$ or from} integral equation theories; see Refs.\ \onlinecite{evans1992density, lutsko, hansen2013theory} and references therein. In this paper we explain why this might not be the best means of testing the accuracy of a free energy functional.

We focus on the much-used excess Helmholtz free energy functional [Eq.\ \eqref{eq:ACE:RPA_DFT} below] that treats the attractive inter-particle interactions in a mean-field (MF) or van der Waals approximation. This {standard} MF DFT is considered the `work-horse' for applications of DFT to interfaces, adsorption and many other inhomogeneous situations, see e.g.\ Refs.\ \onlinecite{evans1992density, lowen2002density, lutsko, hansen2013theory, evans1987phase, stewart2014layering, evans2015local, chacko2017solvent}. Taking two derivatives of this functional and evaluating for a uniform density $\rho_b$, corresponding to the (bulk) fluid, leads to the pair direct correlation function $c^{(2)}_\mathrm{RPA}(r)$ that forms the basis for the well-known random phase approximation (RPA).\cite{evans1979nature, hansen2013theory, likos2001effective,barker1976liquid} Recognizing this connection between the MF DFT and the RPA, one might assume that the fluid structure, thermodynamics and phase behaviour predicted by the MF DFT is of similar quality to that resulting from the RPA closure to the bulk Ornstein-Zernike (OZ) equation. Here, we argue: (i) Results from the MF DFT are more accurate than one might expect from examining bulk pair correlation functions extracted from the RPA closure to the OZ equation and (ii) one should not judge the quality of the MF DFT, or any other approximate DFT, solely upon what the approximate functional generates by taking two functional derivatives. We make these arguments explicit by considering the test-particle limit of Percus,\cite{percus1962approximation} i.e.\ we calculate the inhomogeneous fluid density profile $\rho(r)$ around a fixed particle of the same type as the surrounding fluid. This enables us to calculate the fluid radial distribution function $g(r)=\rho(r)/\rho_b$. By rewriting the Euler-Lagrange equation obtained from minimizing the MF DFT in this test particle limit and comparing with the OZ equation, we identify {additional terms somewhat akin to} a hybrid closure relation {and certainly different from the RPA}. The additional terms {also suggest} that the MF DFT, treated in the test particle limit, is much superior to standard OZ with the RPA closure. We argue that this observation should carry through more generally for arbitrary external potentials, not just in the test-particle limit.

Our paper proceeds as follows: In Sec.~\ref{MFDFT:sec:OZ_all} we describe briefly the RPA in the context of the OZ equation. In Sec.~\ref{sec:MF_DFT} we describe the MF DFT that generates the RPA and indicate why one might expect this DFT to yield much better results for $g(r)$, within the test-particle procedure, than in the standard RPA treatment. In Sec.~\ref{sec:illustration} we illustrate and confirm our arguments by presenting results for $g(x)$ and $S(k)$ for a model one-dimensional (1D) fluid where the pair correlation functions are known exactly and where the MF DFT can be implemented with the exact reference free energy functional, i.e. that pertaining to hard-rods. Finally, in Sec.\ \ref{sec:conclusion} we discuss the general implications of our results. We also explain that {for the same choice of reference system (hard rods) the MF} DFT results are identical to those one would obtain from the local molecular field theory (LMF) of Weeks and co-workers{\cite{weeks1997intermolecular, weeks1998roles, weeks2002connecting}} for this particular model fluid.

\section{Integral equation approach: OZ equation and RPA}
\label{MFDFT:sec:OZ_all}

The two-body pair direct correlation function, $c^{(2)}(r)$, is usually defined via the OZ equation,\cite{hansen2013theory} which for a uniform and isotropic fluid is
\begin{equation}
	h(r) = c^{(2)}(r) + \rho_b \int \dr' c^{(2)}(|\rr-\rr'|) h(r')
	\label{eq:ACE:OZ}
\end{equation}
where $h(r)$ is the total correlation function and $\rho_b$ is the density of the (bulk) fluid. The OZ approach to calculating $h(r)$ is to split the correlations present in $h(r)$ into a direct part, which describes the `direct' correlations acting over a range of order that of the interaction pair potential, and an `indirect' part, i.e.\ the remainder described by the convolution integral. Note that the total correlation function $h(r) = g(r) - 1$, where $g(r)$ is the radial distribution function. Fourier transforming Eq.~\eqref{eq:ACE:OZ} yields an algebraic relation:
\begin{equation}
	\hat{h}(k) = \frac{\hat{c}(k)}{1-\rho_b\hat{c}(k)},
	\label{eq:ACE:OZ_FT}
\end {equation}
where $\hat{h}(k)$ and $\hat{c}(k)$ are the Fourier transforms of $h(r)$ and $c^{(2)}(r)$, respectively. $\hat{h}(k)$ is closely related to the static structure factor\cite{hansen2013theory}
\begin{equation}
	S(k) = 1 + \rho_b\hat{h}(k).
	\label{eq:ACE:s_k_h_hat}
\end{equation}

Consider a fluid composed of particles interacting via the pair potential $v(r)$. To calculate $h(r)$ one must supplement the OZ Eq.~\eqref{eq:ACE:OZ} with a further, closure relation between $c^{(2)}(r)$ and $h(r)$. The exact closure relation is usually expressed as\cite{hansen2013theory}
\begin{equation}
	c^{(2)}(r)= h(r) - \ln(h(r)+1) - \beta v(r) + B(r)
	\label{eq:ACE:exact _closure_relation}
\end{equation}
where $\beta =  (k_B T)^{-1}$, $k_B$ is Boltzmann's constant, $T$ is the temperature and $B(r)$ is termed the bridge function. $B(r)$ is not known exactly for any 3D fluid. In order to make progress approximations must be made. For example, the hypernetted-chain approximation (HNC) corresponds to $B(r) = 0$.\cite{hansen2013theory,barker1976liquid} Formally $B(r)$ is the sum of the bridge or elemental diagrams missing in HNC. Suppose $v(r)$ can be split as follows:
\begin{equation}
	v(r)=v_0(r)+v_1(r),
	\label{eq:ACE:pot_split}
\end{equation}
where $v_0(r)$ is a suitably chosen reference potential, usually the purely repulsive part of $v(r)$. Then the remainder $v_1(r)$ usually incorporates the attractive part of the interaction between particles. The simple closure relation
\begin{equation}
	c_\mathrm{RPA}^{(2)}(r)\equiv c_0^{(2)}(r) -\beta v_1(r),
	\label{eq:ACE:RPA_from_OZ}
\end{equation}
where $c_0^{(2)}(r)$ is the pair direct correlation function for the (purely repulsive) reference system with the same density $\rho_b$, defines the RPA. Note that Eq.\ \eqref{eq:ACE:RPA_from_OZ} enforces the correct asymptotic behaviour: $c^{(2)}(r)\sim-\beta v(r)$, $r\to\infty$ for a fluid away from its critical point. Inserting the Fourier transform of \eqref{eq:ACE:RPA_from_OZ} into \eqref{eq:ACE:OZ_FT} yields the standard RPA result for the structure factor:\cite{hansen2013theory}
\begin{equation}
	S_\mathrm{RPA}(k) = \frac{S_0(k)}{1 + \rho_b\beta\hat{v}_1(k)S_0(k)}
	\label{eq:ACE:S_RPA}
\end{equation}
where $S_0(k)$ is the structure factor of the reference system. The Fourier transform $\hat{v}_1(k)$ is assumed to exist. Often the further approximation $c_0^{(2)}(r)\approx c_\mathrm{HS}^{(2)}(r)$ is made, where $c_\mathrm{HS}^{(2)}(r)$ is the pair direct correlation function for a hard-sphere (HS) fluid at the same density with suitably chosen effective particle diameter $\sigma$.\cite{hansen2013theory, barker1976liquid} Recall that accurate expressions for the reference $c_\mathrm{HS}^{(2)}(r)$ exist and for a fluid of 1D hard-rods $c_\mathrm{HS}^{(2)}(r)$ is known exactly (see below). The RPA closure relation in Eq.~\eqref{eq:ACE:RPA_from_OZ} has been used extensively in the theory of simple and complex liquids.\cite{hansen2013theory, evans1992density, evans1979nature, likos2001effective, lowen2002density} The reliability of the corresponding OZ result for the structure factor \eqref{eq:ACE:S_RPA} depends on the particular model system and the choice of reference potential.

\section{The mean-field DFT and the RPA}
\label{sec:MF_DFT}

\subsection{The non-uniform fluid}

Consider now the fluid composed of particles interacting via the pair potential $v(r)$, split as in Eq.~\eqref{eq:ACE:pot_split} {and subject to an external potential $V(\rr)$. The corresponding one-body density is $\rho(\rr)$}. Suppose too that we have an accurate DFT for the reference system, indicated by the subscript ``0'', i.e.\ with particles interacting via the potential $v_0(r)$. The intrinsic Helmholtz free energy functional can be written as:\cite{evans1992density}
\begin{eqnarray}\label{eq:ACE:F_int}
	F[\rho(\rr)]=&F_0&[\rho(\rr)]\\
	&+&\frac{1}{2}\int_0^1 d\lambda \int \dr \int\dr'\rho_\lambda^{(2)}(\rr,\rr')v_1(|\rr-\rr'|),
	\nonumber
\end{eqnarray}
where $F_0[\rho]$ is the corresponding functional for the reference system. This exact expression is obtained from a thermodynamic integration `turning on' the potential $v_1(r)$ between the particles via the integration parameter $\lambda$. The two-body density distribution function $\rho_\lambda^{(2)}(\rr,\rr')$ is that for the system with interaction potential
\begin{equation}
	v_\lambda(r)=v_0(r)+\lambda v_1(r), \,\,\,\,\, 0\leq\lambda\leq1.
\end{equation}
In deriving Eq.~\eqref{eq:ACE:F_int} one must impose an external potential, varying with $\lambda$, that ensures the equilibrium one-body density remains $\rho(\rr)$ at each value of $\lambda$.\cite{evans1992density} The {standard} MF DFT approximation is obtained by assuming that (i) $\rho_\lambda^{(2)}(\rr,\rr')$ does not change much as $\lambda$ is varied from 0 to 1 and, more drastically, (ii)
\begin{equation}
	\rho_\lambda^{(2)}(\rr,\rr')\approx\rho(\rr)\rho(\rr'), \,\,\,\,\, 0\leq\lambda\leq1.
	\label{eq:ACE:rho2_approx}
\end{equation}
Eq.\ \eqref{eq:ACE:rho2_approx} clearly constitutes a mean-field treatment of the `perturbation' $\lambda v_1(r)$. It follows that the MF DFT approximation for the excess (over ideal) free energy functional\footnote{The correlations neglected in \eqref{eq:ACE:RPA_DFT} are incorporated into the functional $F_{\mathrm{corr}}[\rho]$ defined in Eq.~(3.4.11) of Ref.\ \onlinecite{hansen2013theory}.} is:
\begin{equation}
	F^\mathrm{ex}[\rho(\rr)]\approx F_0^\mathrm{ex}[\rho(\rr)]+\frac{1}{2} \int \dr \int\dr'\rho(\rr)\rho(\rr')v_1(|\rr-\rr'|).
	\label{eq:ACE:RPA_DFT}
\end{equation}
Within DFT\cite{evans1979nature, evans1992density, hansen2013theory} two functional derivatives of $-\beta F^\mathrm{ex}[\rho]$ with respect to the density yields the pair direct correlation function. From Eq.\ \eqref{eq:ACE:RPA_DFT}, and evaluating for a uniform (bulk) fluid $\rho(\rr)=\rho_b$, we obtain the RPA approximation in Eq.~\eqref{eq:ACE:RPA_from_OZ}, since
\begin{equation}
	c_0^{(2)}(|\rr-\rr'|)=-\frac{\delta^2 \beta F_0^\mathrm{ex}[\rho]}{\delta\rho(\rr)\delta\rho(\rr')}\bigg|_{\rho(\rr)=\rho_b}.
	\label{eq:ACE:c2_0}
\end{equation}
The approximations inherent in Eq.\ \eqref{eq:ACE:rho2_approx} imply correlations are omitted and so one must be sceptical about the accuracy of the MF functional \eqref{eq:ACE:RPA_DFT}. {Recognising that taking two functional derivatives of \eqref{eq:ACE:RPA_DFT} yields the RPA (\ref{eq:ACE:RPA_from_OZ}, \ref{eq:ACE:S_RPA}) for bulk correlation functions provides some useful insight into the status of the MF DFT. It is tempting then to argue that employing the MF DFT \eqref{eq:ACE:RPA_DFT} should lead to results with similar accuracy to those obtained from the RPA for bulk liquids. However, this argument is at best misleading. In practical applications of \eqref{eq:ACE:RPA_DFT}, or any other DFT approximation, one works at the one-body level which requires only a single functional derivative. We explain and illustrate this below within the context of the test particle procedure for calculating $g(r)$.}

\subsection{The Percus test particle procedure}

Percus proved\cite{percus1962approximation} that one can determine the radial distribution function $g(r)$ by calculating the density profile $\rho(\rr)=\rho(r)$ around a fixed particle that exerts an external potential $V(\rr)\equiv v(r)$ on the fluid. Then the radial distribution function $g(r)=\rho(r)/\rho_b$. Within DFT, $\rho(r)$ is obtained by minimising the grand potential functional $\Omega[\rho]=F[\rho]-\int\dr[\mu-V(\rr)]\rho(\rr)$, where $\mu$ is the chemical potential. Using \eqref{eq:ACE:RPA_DFT}, the resulting Euler-Lagrange equation is
\begin{widetext}
\begin{eqnarray}
	\frac{\delta\Omega[\rho]}{\delta\rho}=k_BT
	\ln[\Lambda^3\rho(r)]
	+\frac{\delta F_0^\mathrm{ex}[\rho]}{\delta\rho}
	+\int\dr'\rho(r')v_1(|\rr-\rr'|)+v(r)-\mu=0,
	\label{eq:ACE:EL_eq}
\end{eqnarray}
where $\Lambda$ is the (irrelevant) thermal de-Broglie wavelength. For $r\to\infty$, away from the fixed test-particle, the density $\rho(r)\to\rho_b$, so within the approximation Eq.\ \eqref{eq:ACE:RPA_DFT} we obtain the following relation between the chemical potential $\mu$ and the bulk density $\rho_b$:
\begin{equation}
	\mu=k_BT\ln[\Lambda^3\rho_b]
	+\frac{\delta F_0^\mathrm{ex}[\rho]}{\delta\rho}\bigg|_{\rho_b}+\rho_b\int\dr v_1(r).
	\label{eq:ACE:mu}
\end{equation}
We make a functional Taylor expansion about the bulk density:
\begin{eqnarray}
	\frac{\delta F_0^\mathrm{ex}[\rho]}{\delta\rho}=\frac{\delta F_0^\mathrm{ex}[\rho]}{\delta\rho}\bigg|_{\rho_b}
	+\int\dr'(\rho(\rr')-\rho_b)\frac{\delta^2F_0^\mathrm{ex}[\rho]}{\delta\rho(\rr)\delta\rho(\rr')}\bigg|_{\rho_b}
	+H_0[\rho(\rr)],
	\label{eq:ACE:Taylor}
\end{eqnarray}
where $H_0[\rho(\rr)]$ denotes all higher order terms; these are $\sim{\cal O}([\rho-\rho_b]^2)$ and higher. From Eqs.\ \eqref{eq:ACE:c2_0}--\eqref{eq:ACE:Taylor}, we obtain:
\begin{equation}
	0=k_BT\ln\left(\frac{\rho(r)}{\rho_b}\right)+\int\dr'(\rho(r')-\rho_b)\left[-k_BTc_0^{(2)}(|\rr-\rr'|)+v_1(|\rr-\rr'|)\right]+H_0[\rho(r)]+v(r),
\end{equation}
which eliminates $\mu$. Multiplying through by $-\beta$ and adding $(\rho(r)-\rho_b)/\rho_b$ to both sides we obtain:
\begin{align}
	\frac{(\rho(r)-\rho_b)}{\rho_b}=\frac{(\rho(r)-\rho_b)}{\rho_b}-\ln\left(\frac{\rho(r)}{\rho_b}\right)-\beta v(r)-\beta H_0[\rho(r)] 
	+\rho_b\int\dr'\frac{(\rho(r')-\rho_b)}{\rho_b}\left[c_0^{(2)}(|\rr-\rr'|)-\beta v_1(|\rr-\rr'|)\right].
	\label{eq:ACE:OZ-like}
\end{align}
This is the equation for the density profile, equivalent to $g(r)$, in the test particle treatment of Percus, as determined by the MF DFT in Eq.\ \eqref{eq:ACE:RPA_DFT}. If we set $v_1(r)=0$, then we obtain the following equation for the total correlation function $h_0(r)=g_0(r)-1$ of the reference system:
\begin{equation}
	h_0(r)=h_0(r)-\ln\left(h_0(r)+1\right)-\beta v_0(r)-\beta H_0[\rho_bg_0(r)]+\rho_b\int\dr'h_0(r')c_0^{(2)}(|\rr-\rr'|).
\end{equation}
\end{widetext}
\hspace{0cm}{Suppose we know the exact functional $F_0[\rho]$, and therefore $c_0^{(2)}(r)$, then comparison with Eq.\ \eqref{eq:ACE:exact _closure_relation} {and use of the OZ equation \eqref{eq:ACE:OZ}} allows us to identify $-\beta H_0[\rho_bg_0(r)]$ as the exact} bridge-function $B_0(r)$ of the reference system.

Returning to the full system, we see that Eq.\ \eqref{eq:ACE:OZ-like} is an Ornstein-Zernike-like equation [see Eq.\ \eqref{eq:ACE:OZ}] with the RPA closure \eqref{eq:ACE:RPA_from_OZ} for the pair direct correlation function $c^{(2)}(r)$ inside the convolution integral, but with a different {closure approximation for $c^{(2)}(r)$ [see Eq.\ \eqref{eq:ACE:exact _closure_relation}] appearing outside. Specifically, the sum of the first four terms on the right-hand side of \eqref{eq:ACE:OZ-like} correspond formally to the exact expression for $c^{(2)}(r)$ but with the bridge function $B(r)$ replaced by $B_0(r)$.

The form of Eq.~\eqref{eq:ACE:OZ-like}, which follows from the standard MF DFT functional \eqref{eq:ACE:RPA_DFT}, suggests that calculating the radial distribution function $g(r)$ via the test-particle route might yield results better than those given by $g_\mathrm{RPA}(r)$, obtained by solving the OZ equation together with the RPA closure \eqref{eq:ACE:RPA_from_OZ}, i.e.\ by Fourier inverting the RPA structure factor \eqref{eq:ACE:S_RPA}. This is evident for a fluid in which the pair potential $v(r)$ has a hard-core of diameter $\sigma$ since solving the Euler-Lagrange equation \eqref{eq:ACE:EL_eq} guarantees the exact core condition $\rho(r) = 0$, $r < \sigma$ is satisfied which is, of course, not the case for $g_\mathrm{RPA}(r)$, given by \eqref{eq:ACE:S_RPA}. In the latter $g_\mathrm{RPA}(r)$ is not identically zero inside the hard-core. We note that the application of any reasonable non-local DFT in the test particle procedure enforces the core condition. In the footnote\footnote{There are, of course, other theories that enforce the hard-core condition on $g(r)$. The well-known Mean-Spherical Approximation (MSA)\cite{hansen2013theory, lebowitz1966mean} enforces this and sets $c^{(2)}(r) = -\beta v(r)$ outside the core. Perhaps more pertinent to our present discussion is the Optimized RPA (ORPA).\cite{hansen2013theory, andersen1972roles} The ORPA invokes the RPA closure \eqref{eq:ACE:RPA_from_OZ} but, in addition, seeks to vary the perturbation potential $v_1(r)$ inside the hard core, where this is not uniquely defined, so that $g(r)=0$. This constraint is equivalent to requiring the functional derivative of the RPA free energy with respect to $v_1(r)$ to be zero inside the hard core. In our present approach, there is no attempt to minimize the free energy with respect to the perturbation potential, so for $r>\sigma$, $g(r)$ depends weakly on the choice of $v_1(r)$ inside the core. There is no reason to expect the MF DFT to perform as well as the ORPA -- at least for a Lennard-Jones type fluid at high densities where the ORPA is known to be very accurate.\cite{hansen2013theory, andersen1972roles}
 
For completeness, we also mention the Reference HNC (RHNC), see e.g.\ Ref.~\onlinecite{hansen2013theory}, which improves upon the HNC integral equation by approximating the bridge function by that of a reference system, usually hard-spheres. Although the integral equation \eqref{eq:ACE:OZ-like} that emerges from our MF DFT shares a feature of the RHNC, in that $B(r)$ is replaced by $B_0(r)$, this is implemented only in one part of the right hand side. There is no reason to expect the MF DFT to be as accurate as the RHNC.} we mention briefly relationships to other theories of liquids.

This observation concerning the test-particle procedure has repercussions for more general} external potentials $V(\rr)$. Solving the corresponding Euler-Lagrange equation, based on the seemingly crude MF functional \eqref{eq:ACE:RPA_DFT}, yields equilibrium density profiles $\rho(\rr)$ that are often very accurate -- see Refs.\ \onlinecite{evans1992density, hansen2013theory} and references therein. Investigation of hard/impenetrable potentials is once again illuminating. For a planar hard wall, with $V(z) =\infty$ for $z<0$, the density profile satisfies $\rho(z) =0$, $z<0$. Moreover, for a sensible choice of a (non-local) DFT for the reference system the profile will satisfy\cite{evans1992density} the wall contact sum-rule: $\beta\rho(0^+)=p(\rho_b)$. The right hand side is the pressure of the bulk fluid, far from the hard wall, obtained from the bulk free energy $F[\rho_b]$ with $\rho_b=\rho(\infty)$.

In the following Sec.\ \ref{sec:illustration} we compare the results for $g_\mathrm{RPA}(x)$ and $S_\mathrm{RPA}(k)$ {\eqref{eq:ACE:S_RPA}} with those obtained from {standard} MF DFT \eqref{eq:ACE:RPA_DFT}, implemented within the Percus test particle prescription for a 1D fluid. The reference system is the hard-rod fluid for which the free energy functional is known exactly, and therefore its correlation functions and thermodynamics. Moreover, the pair correlation functions and the thermodynamics of the full system for uniform (bulk) densities are also known exactly. By considering this model fluid we can make a stringent examination of some of the basic approximations employed in classical DFT.

\section{Illustrative results for a 1D fluid}
\label{sec:illustration}

\begin{figure*}[!t]
  \includegraphics[width=0.9\textwidth]{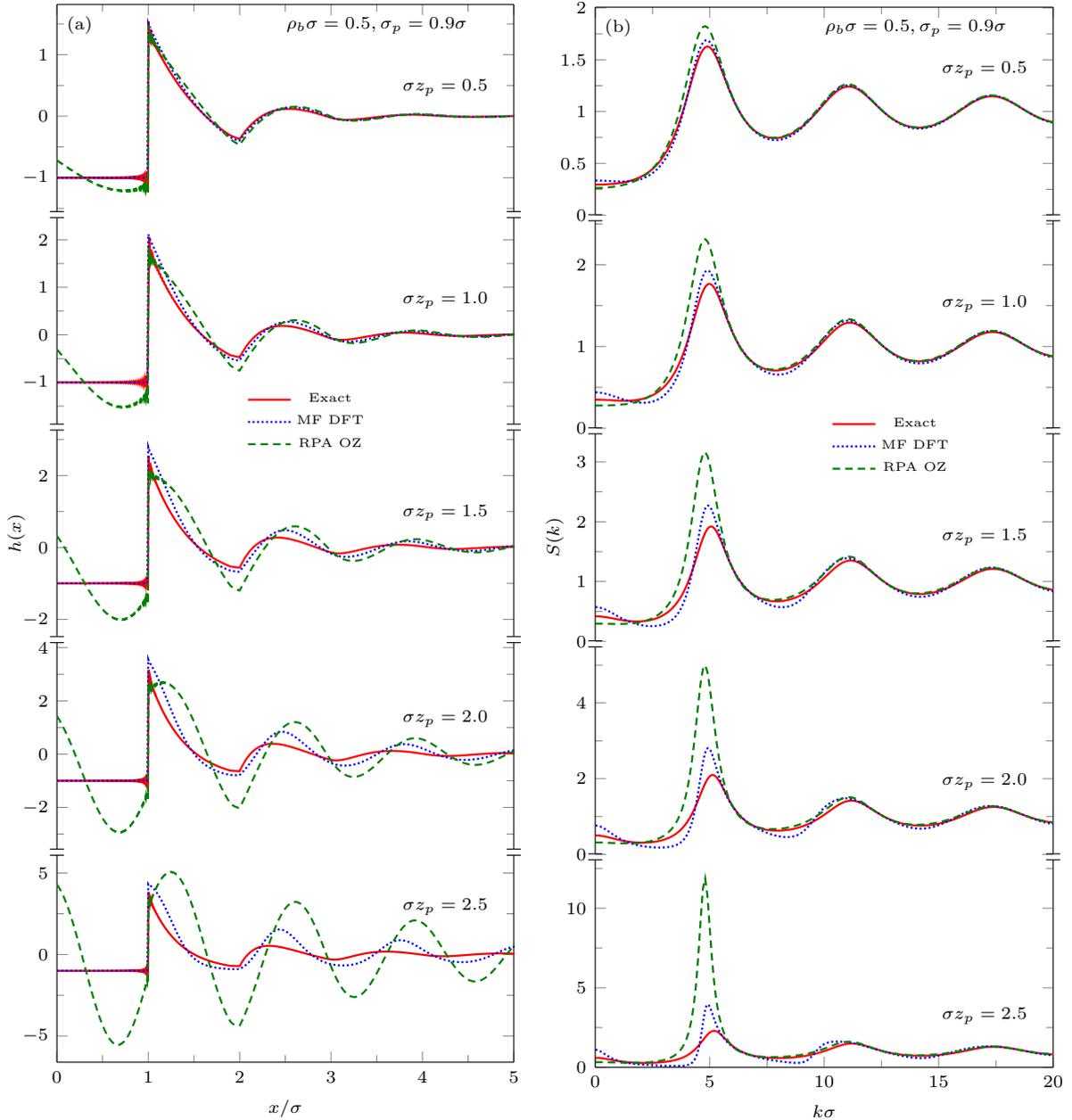}
	\caption[The total correlation function $h(x)$ and the structure factor $S(k)$ for hard-rods for various values of attraction strength $z_p$ ]{(a) The total correlation function $h(x)$ and (b) the static structure factor $S(k)$ for hard-rods with an attractive tail potential \eqref{eq:ACE:tail} for various values of the attraction strength $z_p$ with fixed $\sigma_p = 0.9\sigma$ and $\rho_b\sigma = 0.5$. We compare the exact results (solid line) with $g_\mathrm{RPA}(x)-1$ and $S_\mathrm{RPA}(k)$ from the RPA closure \eqref{eq:ACE:S_RPA} to the OZ equation (dashed) and with those from the MF DFT~\eqref{eq:ACE:RPA_DFT} using the test particle route (dotted). Note the different scales on each y-axis.}
	\label{fig:rho50sp90zp}
\end{figure*}

\begin{figure*}[!t]
  \includegraphics[width=0.9\textwidth]{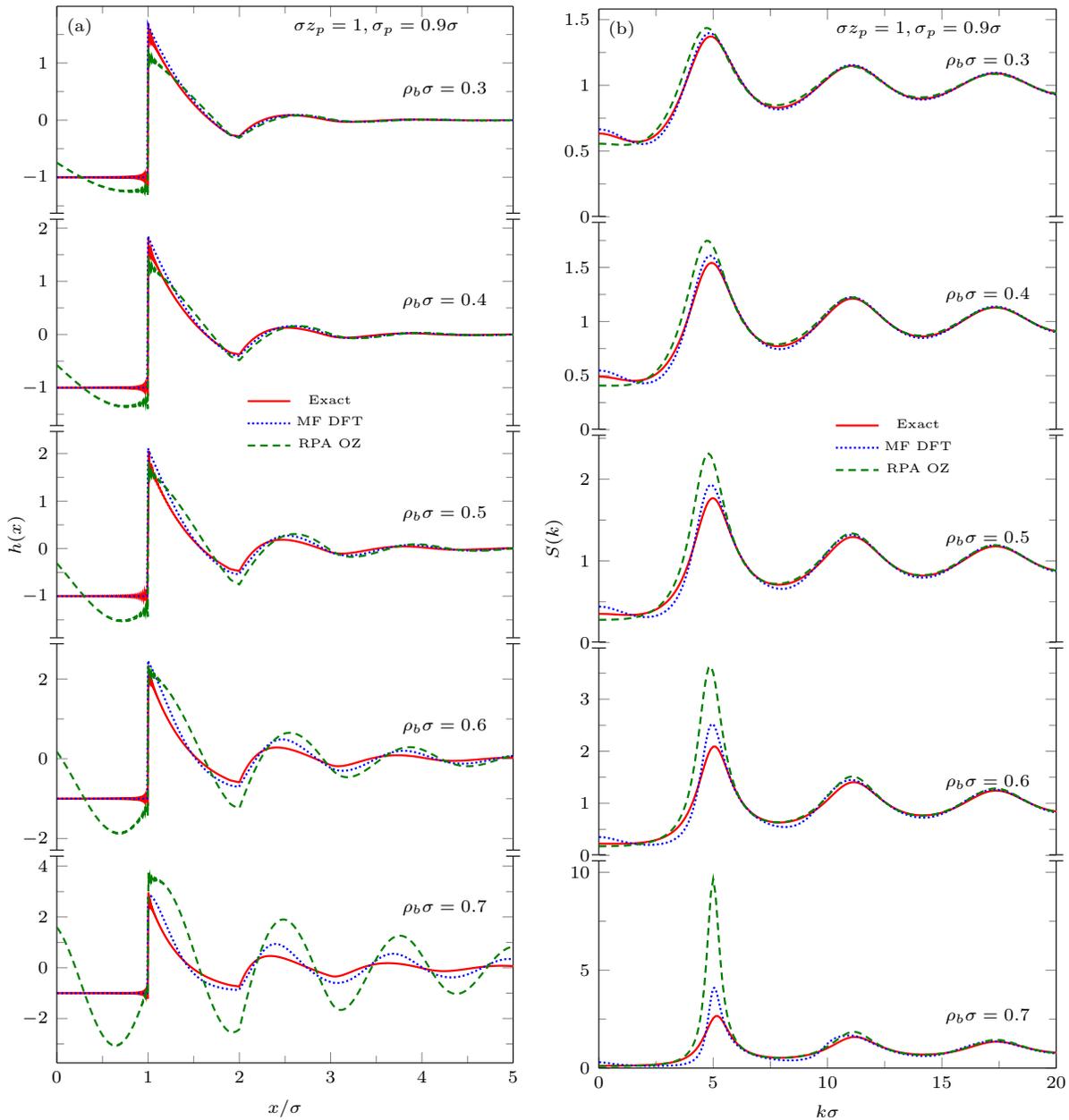}
	\caption[The total correlation function $h(x)$ and the structure factor $S(k)$ for hard-rods with tail potential \eqref{eq:ACE:tail} for various values of bulk density $\rho_b$]{(a) The total correlation function $h(x)$ and (b) the static structure factor $S(k)$ for hard-rods with an attractive tail potential \eqref{eq:ACE:tail} for various values of bulk density $\rho_b$ with fixed $\sigma z_p = 1$ and $\sigma_p = 0.9\sigma$. The key is the same as in Fig.\ \ref{fig:rho50sp90zp}.}
	\label{fig:sp90zp10rho}
\end{figure*}

\begin{figure*}[!t]
     \includegraphics[width=0.9\textwidth]{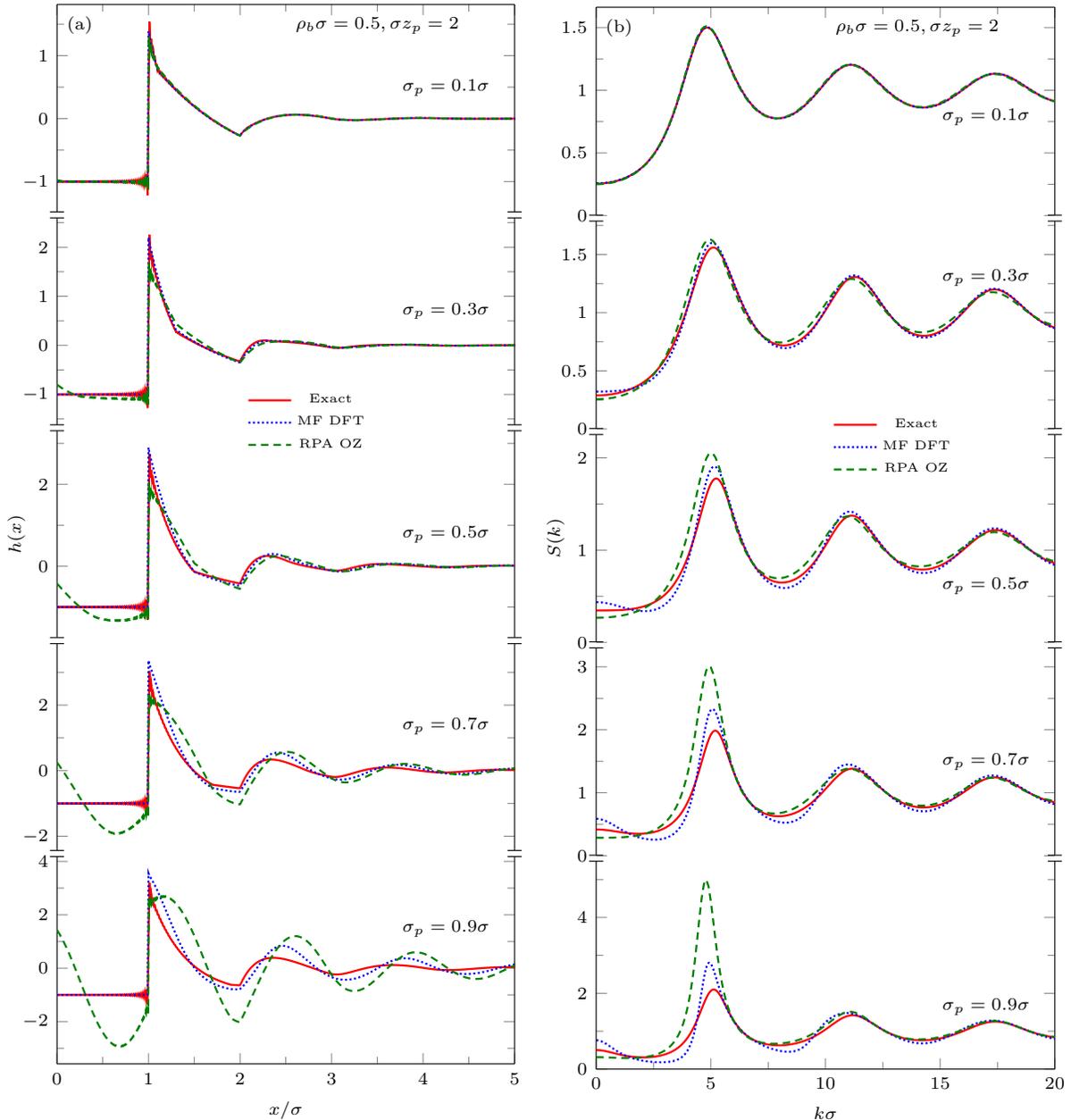}
	\caption[The total correlation function $h(x)$ and the structure factor $S(k)$ for hard-rods for various values of attraction range $\sigma_p$ ]{{(a) The total correlation function $h(x)$ and (b) the static structure factor $S(k)$ for hard-rods with an attractive tail potential \eqref{eq:ACE:tail} for various values of the attraction range $\sigma_p$ with fixed $\sigma z_p = 2$ and $\rho_b\sigma=0.5$. The key is the same as in Fig.\ \ref{fig:rho50sp90zp}.}}
	\label{fig:sigma_p}
\end{figure*}

In order to illustrate and support the observations made above, we consider a system of 1D hard-particles (rods on a line) with an additional attractive interaction between pairs of neighbouring rods. As mentioned above, this 1D system is chosen because we know the ingredients of the theory exactly and can therefore test carefully the accuracy of the various approximations. For a fluid of 1D rods with only nearest neighbour interactions $v(x)$, one finds the following exact expression for the structure factor\cite{percus1982one}
\begin{equation}
	{S(k)}= \frac{  1 - e^{ -\beta [ \mu(p+ik/\beta) - 2\mu(p) +  \mu(p-ik/\beta) ] } }
		{ ( 1 - e^{  -\beta[ \mu(p+ik/\beta) - \mu(p) ] }) ( 1 - e^{  -\beta[ \mu(p-ik/\beta) - \mu(p) ] } ) }
\end{equation}
where $p$ is the pressure and $\mu(p)$ is the chemical potential which are known exactly.\cite{takahashi1942simple,brader2002exactly} The above equation along with Eqs.~\eqref{eq:ACE:OZ} -- \eqref{eq:ACE:s_k_h_hat} can be used to obtain the distribution functions of the uniform 1D system.

We consider 1D rods on a line interacting via the pair potential $v(x)=v_0(x)+v_1(x)$, where $x$ is the distance between the centres of a pair of particles -- c.f.\ Eq.\ \eqref{eq:ACE:pot_split}. $v_0(x)$ is the hard-rod potential
\begin{equation}
    v_0(x) = \left\{
           \begin{array}{l l}
             \infty
               & \quad  {|x| \le \sigma}
             \\
             0
               & \quad {|x| > \sigma}
           \end{array}
         \right.
       \label{eq:ACE:pot}
\end{equation}
and the attractive tail potential (also considered in Ref.~\onlinecite{archer2013relationship}) is:
\begin{equation}
    \beta v_1(x) = \left\{
           \begin{array}{l l}
             0
               & \quad  {|x| \le \sigma}
             \\
             -z_p(\sigma +\sigma_p - |x| )
               & \quad {\sigma < |x| \le \sigma+\sigma_p}
             \\
             0
               & \quad {|x| > \sigma+\sigma_p}.               
           \end{array}
         \right.
       \label{eq:ACE:tail}
\end{equation}
The above potential is the 1D analogue of the Asakura-Oosawa potential for the effective colloid-colloid potential of hard-sphere colloids of diameter (length) $\sigma$ mixed with ideal polymers: $\sigma_p$ is the `length' of polymer coils and $z_p$ is the fugacity of ideal polymers.\cite{dijkstra1999phase,archer2013relationship,brader2002exactly} 

{For the hard-rods reference system ``0'' Percus\cite{percus1976equilibrium, percus1982one} derived an exact expression for the excess Helmholtz free energy functional
\begin{equation}
	  F_0^\mathrm{ex}[\rho]\equiv F_\mathrm{HR}^\mathrm{ex}[ \rho ]  = - \int \rho(x) \ln\left[ 1 - t(x) \right]  \mathrm{d}x,
	  \label{eq:ACE:exact_excess_percus}
\end{equation}
where the weighted density is
\begin{equation}
	t(x) = \int_{x-\sigma}^x \rho(x') \mathrm{d}x'.
\end{equation}
Taking the functional derivative of \eqref{eq:ACE:exact_excess_percus} and adding the contribution from the ideal-gas free energy $F^\mathrm{id}[\rho]$, we obtain
\begin{equation}
	\frac{\delta F_0[\rho]}{\delta\rho} = \ln \frac{ \Lambda\rho(x)}{1-t(x)} +  \int_x^{x+\sigma} \frac{\rho(x')}{1-t(x')} \mathrm{d}x'.
	  \label{eq:ACE:exact_func_deriv}
\end{equation}
By differentiating again and using \eqref{eq:ACE:c2_0} the exact direct correlation function of a uniform  fluid of hard-rods of length $\sigma$ and density $\rho_b$ is derived:
\begin{equation}
	c^{(2)}_\mathrm{HR}(|x-x'|) = -\Theta(\sigma-|x-x'|)\frac{1-\rho_b|x-x'|} {(1-\rho_b\sigma)^2},
	\label{eq:ACE:Exact_c_HR}
\end{equation} 
where $\Theta$ is the Heaviside step function. This can be used to construct the RPA approximation \eqref{eq:ACE:RPA_from_OZ} for the 1D system.}

Comparison of the exact solution (solid line) with DFT results [calculated using the test particle equation~\eqref{eq:ACE:EL_eq} with \eqref{eq:ACE:exact_func_deriv} and $v(x)$ (dotted)] and results for $g_\mathrm{RPA}(x)$ and $S_\mathrm{RPA}(k)$ from the RPA approximation to the OZ equation \eqref{eq:ACE:S_RPA} {with $v_1(x)$ (dashed)}, is shown in Figs.~\ref{fig:rho50sp90zp}, \ref{fig:sp90zp10rho} and \ref{fig:sigma_p}.{\footnote{Note that for $\lambda =1$, $S_{\mathrm{MF}\lambda}(k)$, introduced in Eq.~(41) of Ref.~\onlinecite{archer2013relationship}, is identical to $S_\mathrm{RPA}(k)$ defined here.}} {In Fig.~\ref{fig:rho50sp90zp} we fix the bulk density $\rho_b\sigma = 0.5$ and the attraction range $\sigma_p=0.9\sigma$ while varying the attraction strength parameter $z_p$. In Fig.~\ref{fig:sp90zp10rho} we fix the attraction strength $\sigma z_p =1$ and range $\sigma_p=0.9\sigma$, while varying the bulk density $\rho_b\sigma$. Finally, in Fig.~\ref{fig:sigma_p} we fix the attraction strength $\sigma z_p =2$ and the bulk density $\rho_b\sigma=0.5$ while varying the attraction range $\sigma_p$}.

As expected, $g_\mathrm{RPA}(x)$ fails to satisfy the core condition $g(x) = 0$ when $x<\sigma$, whereas the DFT enforces this. Outside the core of the hard-rod, both the DFT and RPA differ from the exact solution. However, the DFT results for $g(x)$ are much closer to the exact solution. As we increase the attraction strength or the density, both the DFT and RPA results deviate increasingly from the exact solution. $g_\mathrm{RPA}(x)$ displays only weakly damped oscillations. The same type of pattern is also observed in the sequence of structure factors displayed in Figs.~\ref{fig:rho50sp90zp}(b) and \ref{fig:sp90zp10rho}(b). The RPA greatly overestimates the height of the first peak in $S(k)$ as $z_p\sigma$ (attractive well-depth) or $\rho_b\sigma$ are increased. It is important to note that when these parameters are increased beyond the values considered here, $S_\mathrm{RPA}(k)$ diverges at the first peak.\cite{archer2013relationship} {Note that for a given choice of pair potential the number density at which the divergence occurs is identical within both the RPA and MF DFT since the linearized form of both theories is identical; only the terms that are non-linear in $h(x)$ are different. However, the height of the first peak in the structure factor obtained from the MF DFT is always lower and closer to the exact value. It is also noteworthy that both the RPA and MF DFT capture well the behaviour in $S(k)$} at larger $k\sigma$.

{The results in Fig.\ \ref{fig:sigma_p} show that for fixed attraction strength $z_p$ both the MF DFT and the RPA become less accurate as the range of the potential $\sigma_p$ increases. In particular the height of the first peak in $S(k)$ is overestimated. It appears that the integrated strength of the attraction, measured by the product $z_p\sigma_p$, is important in determining under what conditions both approximations are accurate.}

\section{Discussion}
\label{sec:conclusion}

In order to assess the physical content, and hence the usefulness, of an approximation for the excess free energy functional, it is not sufficient to take two functional derivatives, compute the bulk fluid pair direct correlation function $c^{(2)}(r)$, and then enquire how well this quantity performs when used to calculate the pair correlation function $h(r)$ via the OZ route, Eqs.\ \eqref{eq:ACE:OZ}-\eqref{eq:ACE:s_k_h_hat}. In most practical applications of DFT one is concerned with solving the Euler-Lagrange equation for the one-body density profiles and calculating the associated thermodynamic quantities (free energies) that result from minimizing a given (approximate) grand potential functional. This requires taking only a single functional derivative--not two. We have illustrated this point of view by focusing on the standard MF DFT, defined by \eqref{eq:ACE:RPA_DFT}, which treats the attractive part of the inter-particle potential via a simple MF approximation. Working in 1D and computing the density profile induced by a test particle exerting the potential (\ref{eq:ACE:pot}, \ref{eq:ACE:tail}), we determined $g(x)$ and $S(k)$ from the MF DFT. These were compared with the exact results and with those obtained using \eqref{eq:ACE:RPA_from_OZ} with the OZ equation. The latter corresponds to the usual RPA, given by \eqref{eq:ACE:S_RPA}. That the test particle route yields more accurate structure, within a DFT treatment, is not surprising. What is significant is that the MF DFT applied in the test particle situation performs much better for strong inter-particle attraction and for higher densities than the standard RPA. This is evident in Figs.\ \ref{fig:rho50sp90zp}--\ref{fig:sigma_p}.

That it is generally more appropriate to assess the performance of an approximate functional at the one-body rather than at the two-body level is known to the DFT community. The former requires only a single functional derivative with respect to density whereas the latter requires two. Naturally errors build up as further derivatives are taken. Here we are concentrating upon the efficacy of the particular functional \eqref{eq:ACE:RPA_DFT}. Why might the results, from what appears to be a crude approximation, be much better at the one-body density and free energy level than one might expect? We give three separate arguments: (i) as indicated in Sec.\ \ref{sec:MF_DFT}, the Euler-Lagrange equation \eqref{eq:ACE:OZ-like} for $\rho(r)$, {when viewed as an integral equation for $g(r)$}, implies a more sophisticated closure approximation than the RPA \eqref{eq:ACE:RPA_from_OZ} implemented directly in the OZ equation \eqref{eq:ACE:OZ_FT}, which  leads to the RPA expression \eqref{eq:ACE:S_RPA}. (ii) Oettel \cite{oettel2005integral} discusses \eqref{eq:ACE:RPA_DFT} in the context of a powerful and rather general reference functional approach for constructing approximate free energy functionals. By invoking the assumption that the bridge functional for the full system is well-approximated by that of the reference system and by considering expansions about the bulk density he argues that the reference functional approach predicts roughly MF behaviour for the density deviations (from bulk). He concludes that for adsorption problems, such as wetting and drying, a description based on the MF DFT \eqref{eq:ACE:RPA_DFT}, with an accurate reference functional $F_0[\rho]$, should capture all the essential physical features. Oettel\cite{oettel2005integral} also emphasizes that \eqref{eq:ACE:RPA_DFT} has the advantage, over the more sophisticated reference functional approach, of satisfying identically the Gibbs adsorption equation and the wall-contact sum rule. (iii) Weeks and co-workers\cite{weeks1997intermolecular, weeks1998roles, weeks2002connecting} introduced a local molecular field theory (LMF) that has proved to be highly successful in describing the structure and thermodynamics of a variety of non-uniform liquids. The derivation\cite{weeks1998roles, rodgers2008local} of the LMF equation for the effective reference field $\phi_R(\rr)$ starts with the Yvon-Born-Green equation and uses insightful arguments about the form of the conditional singlet densities $\rho(\rr|\rr')$ in the full and reference (mimic) systems. It does not employ concepts from DFT. Although LMF operates at the one-body level, like MF DFT, at first sight there does not appear to be a direct connection between the two approaches. This is not the case. Archer and Evans\cite{archer2013relationship} showed that the LMF equation follows directly from the standard mean-field treatment of attractive interactions as embodied in MF DFT \eqref{eq:ACE:RPA_DFT} and if one has access to the exact functional $F_0[\rho]$ for the same reference system the two theories are equivalent. We note that the derivation of the LMF equation and the relation to DFT is also discussed in the Supporting Information in a recent paper on solvation free energies.\cite{remsing2016long} In the 1D system described in Sec.\ \ref{sec:illustration} the free energy functional of the reference (hard-rod) fluid is known exactly; it is given by the Percus result \eqref{eq:ACE:exact_excess_percus}. It follows that our present results for $g(x)$ obtained using MF DFT and the test particle route are identical to those that would emerge from LMF using hard-rods as the reference system. More generally, in three dimensions, LMF with a hard-sphere reference system would lead to the same $g(r)$ as MF DFT using the test particle route and a very accurate hard-sphere functional for $F_0[\rho]$. Given the success of LMF for a wide variety of fluids, one might argue, albeit circuitously, that the physical arguments and plausible approximations that lead to LMF\cite{weeks1998roles, rodgers2008local, remsing2016long} provide an alternative justification as to why MF DFT might perform better than one might expect. {Of course, there is a caveat. The justification for LMF relies upon the judicious choice of reference fluid, described by $v_ 0(r)$, so that the LMF equation used to treat $v_ 1(r)$, the longer ranged part of the pair potential, captures the essential physics for a given model fluid.\cite{rodgers2008local, remsing2016long} Although the MF DFT treatment of $v_1(r)$ is formally equivalent to that of the LMF, the limitation and drawback of the former is finding an accurate free energy functional for the reference fluid.\cite{archer2013relationship}}

As a final note of caution on assessing the quality of a DFT on the basis of what two derivatives of the free energy functional yields for the (bulk) pair direct correlation function, one should also recall the following functional:
\begin{eqnarray}\label{eq:ACE:RY_DFT}
	F^\mathrm{ex}[\rho(\rr)]\approx F^\mathrm{ex}[\rho_b]+\mu_\mathrm{ex}\int \dr(\rho(\rr)-\rho_b)\hspace{2cm}\\
	-\frac{1}{2 \beta} \int \dr \int\dr'(\rho(\rr)-\rho_b)(\rho(\rr')-\rho_b)c(|\rr-\rr'|).
\nonumber
\end{eqnarray}
This is the well-known Ramakrishnan-Yousouf functional,\cite{ramakrishnan1979first} constructed to yield the `exact' pair direct correlation function, $c(r)=c^{(2)}_\mathrm{exact}(r)$, with $c^{(2)}_\mathrm{exact}(r)$ calculated at the relevant bulk density $\rho_b$.{\footnote{Note that minimizing the grand potential corresponding to \eqref{eq:ACE:RY_DFT} within the test particle procedure leads to the HNC closure for the uniform fluid, provided the direct correlation function is determined self-consistently using the OZ equation\cite{evans1992density, hansen2013theory}}} However, {the functional \eqref{eq:ACE:RY_DFT}} has significant weaknesses. For example, it is unable to describe wetting or drying phenomena\cite{evans1992density, evans1983failure} at substrates nor critical adsorption\cite{evans1986comment} owing to the fact that it is only quadratic in the density deviation.

\section*{Acknowledgements}
B.C. is supported by an EPSRC studentship and R.E. by a Leverhulme Emeritus Fellowship: EM-2016-031.


\begin{thebibliography}{33}%
\makeatletter
\providecommand \@ifxundefined [1]{%
 \@ifx{#1\undefined}
}%
\providecommand \@ifnum [1]{%
 \ifnum #1\expandafter \@firstoftwo
 \else \expandafter \@secondoftwo
 \fi
}%
\providecommand \@ifx [1]{%
 \ifx #1\expandafter \@firstoftwo
 \else \expandafter \@secondoftwo
 \fi
}%
\providecommand \natexlab [1]{#1}%
\providecommand \enquote  [1]{``#1''}%
\providecommand \bibnamefont  [1]{#1}%
\providecommand \bibfnamefont [1]{#1}%
\providecommand \citenamefont [1]{#1}%
\providecommand \href@noop [0]{\@secondoftwo}%
\providecommand \href [0]{\begingroup \@sanitize@url \@href}%
\providecommand \@href[1]{\@@startlink{#1}\@@href}%
\providecommand \@@href[1]{\endgroup#1\@@endlink}%
\providecommand \@sanitize@url [0]{\catcode `\\12\catcode `\$12\catcode
  `\&12\catcode `\#12\catcode `\^12\catcode `\_12\catcode `\%12\relax}%
\providecommand \@@startlink[1]{}%
\providecommand \@@endlink[0]{}%
\providecommand \url  [0]{\begingroup\@sanitize@url \@url }%
\providecommand \@url [1]{\endgroup\@href {#1}{\urlprefix }}%
\providecommand \urlprefix  [0]{URL }%
\providecommand \Eprint [0]{\href }%
\providecommand \doibase [0]{http://dx.doi.org/}%
\providecommand \selectlanguage [0]{\@gobble}%
\providecommand \bibinfo  [0]{\@secondoftwo}%
\providecommand \bibfield  [0]{\@secondoftwo}%
\providecommand \translation [1]{[#1]}%
\providecommand \BibitemOpen [0]{}%
\providecommand \bibitemStop [0]{}%
\providecommand \bibitemNoStop [0]{.\EOS\space}%
\providecommand \EOS [0]{\spacefactor3000\relax}%
\providecommand \BibitemShut  [1]{\csname bibitem#1\endcsname}%
\let\auto@bib@innerbib\@empty
\bibitem [{\citenamefont {Evans}(1979)}]{evans1979nature}%
  \BibitemOpen
  \bibfield  {author} {\bibinfo {author} {\bibfnamefont {R.}~\bibnamefont
  {Evans}},\ }\href@noop {} {\bibfield  {journal} {\bibinfo  {journal} {Adv.
  Phys.}\ }\textbf {\bibinfo {volume} {28}},\ \bibinfo {pages} {143} (\bibinfo
  {year} {1979})}\BibitemShut {NoStop}%
\bibitem [{\citenamefont {Evans}(1992)}]{evans1992density}%
  \BibitemOpen
  \bibfield  {author} {\bibinfo {author} {\bibfnamefont {R.}~\bibnamefont
  {Evans}},\ }in\ \href@noop {} {\emph {\bibinfo {booktitle} {Fundamentals of
  Inhomogeneous Fluids}}},\ \bibinfo {editor} {edited by\ \bibinfo {editor}
  {\bibfnamefont {D.}~\bibnamefont {Henderson}}}\ (\bibinfo  {publisher}
  {Marcel Dekker},\ \bibinfo {address} {New York},\ \bibinfo {year} {1992})\
  Chap.~\bibinfo {chapter} {3}, pp.\ \bibinfo {pages} {85--176}\BibitemShut
  {NoStop}%
\bibitem [{\citenamefont {L{\"o}wen}(2002)}]{lowen2002density}%
  \BibitemOpen
  \bibfield  {author} {\bibinfo {author} {\bibfnamefont {H.}~\bibnamefont
  {L{\"o}wen}},\ }\href@noop {} {\bibfield  {journal} {\bibinfo  {journal} {J.
  Phys.: Condens. Matter}\ }\textbf {\bibinfo {volume} {14}},\ \bibinfo {pages}
  {11897} (\bibinfo {year} {2002})}\BibitemShut {NoStop}%
\bibitem [{\citenamefont {Lutsko}(2010)}]{lutsko}%
  \BibitemOpen
  \bibfield  {author} {\bibinfo {author} {\bibfnamefont {J.~F.}\ \bibnamefont
  {Lutsko}},\ }\href@noop {} {\bibfield  {journal} {\bibinfo  {journal} {Adv.
  Chem. Phys.}\ }\textbf {\bibinfo {volume} {144}},\ \bibinfo {pages} {1}
  (\bibinfo {year} {2010})}\BibitemShut {NoStop}%
\bibitem [{\citenamefont {Hansen}\ and\ \citenamefont
  {McDonald}(2013)}]{hansen2013theory}%
  \BibitemOpen
  \bibfield  {author} {\bibinfo {author} {\bibfnamefont {J.-P.}\ \bibnamefont
  {Hansen}}\ and\ \bibinfo {author} {\bibfnamefont {I.~R.}\ \bibnamefont
  {McDonald}},\ }\href@noop {} {\emph {\bibinfo {title} {Theory of Simple
  Liquids: With Applications to Soft Matter, 4th ed.}}}\ (\bibinfo  {publisher}
  {Elsevier},\ \bibinfo {year} {2013})\BibitemShut {NoStop}%
\bibitem [{\citenamefont {Evans}\ and\ \citenamefont
  {Marconi}(1987)}]{evans1987phase}%
  \BibitemOpen
  \bibfield  {author} {\bibinfo {author} {\bibfnamefont {R.}~\bibnamefont
  {Evans}}\ and\ \bibinfo {author} {\bibfnamefont {U.~M.~B.}\ \bibnamefont
  {Marconi}},\ }\href@noop {} {\bibfield  {journal} {\bibinfo  {journal} {J.
  Chem. Phys.}\ }\textbf {\bibinfo {volume} {86}},\ \bibinfo {pages} {7138}
  (\bibinfo {year} {1987})}\BibitemShut {NoStop}%
\bibitem [{\citenamefont {Stewart}\ and\ \citenamefont
  {Evans}(2014)}]{stewart2014layering}%
  \BibitemOpen
  \bibfield  {author} {\bibinfo {author} {\bibfnamefont {M.~C.}\ \bibnamefont
  {Stewart}}\ and\ \bibinfo {author} {\bibfnamefont {R.}~\bibnamefont
  {Evans}},\ }\href@noop {} {\bibfield  {journal} {\bibinfo  {journal} {J.
  Chem. Phys.}\ }\textbf {\bibinfo {volume} {140}},\ \bibinfo {pages} {134704}
  (\bibinfo {year} {2014})}\BibitemShut {NoStop}%
\bibitem [{\citenamefont {Evans}\ and\ \citenamefont
  {Stewart}(2015)}]{evans2015local}%
  \BibitemOpen
  \bibfield  {author} {\bibinfo {author} {\bibfnamefont {R.}~\bibnamefont
  {Evans}}\ and\ \bibinfo {author} {\bibfnamefont {M.~C.}\ \bibnamefont
  {Stewart}},\ }\href@noop {} {\bibfield  {journal} {\bibinfo  {journal} {J.
  Phys.: Condens. Matter}\ }\textbf {\bibinfo {volume} {27}},\ \bibinfo {pages}
  {194111} (\bibinfo {year} {2015})}\BibitemShut {NoStop}%
\bibitem [{\citenamefont {Chacko}\ \emph {et~al.}(2017)\citenamefont {Chacko},
  \citenamefont {Evans},\ and\ \citenamefont {Archer}}]{chacko2017solvent}%
  \BibitemOpen
  \bibfield  {author} {\bibinfo {author} {\bibfnamefont {B.}~\bibnamefont
  {Chacko}}, \bibinfo {author} {\bibfnamefont {R.}~\bibnamefont {Evans}}, \
  and\ \bibinfo {author} {\bibfnamefont {A.~J.}\ \bibnamefont {Archer}},\
  }\href@noop {} {\bibfield  {journal} {\bibinfo  {journal} {J. Chem. Phys.}\
  }\textbf {\bibinfo {volume} {146}},\ \bibinfo {pages} {124703} (\bibinfo
  {year} {2017})}\BibitemShut {NoStop}%
\bibitem [{\citenamefont {Likos}(2001)}]{likos2001effective}%
  \BibitemOpen
  \bibfield  {author} {\bibinfo {author} {\bibfnamefont {C.~N.}\ \bibnamefont
  {Likos}},\ }\href@noop {} {\bibfield  {journal} {\bibinfo  {journal} {Phys.
  Rep.}\ }\textbf {\bibinfo {volume} {348}},\ \bibinfo {pages} {267} (\bibinfo
  {year} {2001})}\BibitemShut {NoStop}%
\bibitem [{\citenamefont {Barker}\ and\ \citenamefont
  {Henderson}(1976)}]{barker1976liquid}%
  \BibitemOpen
  \bibfield  {author} {\bibinfo {author} {\bibfnamefont {J.~A.}\ \bibnamefont
  {Barker}}\ and\ \bibinfo {author} {\bibfnamefont {D.}~\bibnamefont
  {Henderson}},\ }\href@noop {} {\bibfield  {journal} {\bibinfo  {journal}
  {Rev. Mod. Phys.}\ }\textbf {\bibinfo {volume} {48}},\ \bibinfo {pages} {587}
  (\bibinfo {year} {1976})}\BibitemShut {NoStop}%
\bibitem [{\citenamefont {Percus}(1962)}]{percus1962approximation}%
  \BibitemOpen
  \bibfield  {author} {\bibinfo {author} {\bibfnamefont {J.~K.}\ \bibnamefont
  {Percus}},\ }\href@noop {} {\bibfield  {journal} {\bibinfo  {journal} {Phys.
  Rev. Lett.}\ }\textbf {\bibinfo {volume} {8}},\ \bibinfo {pages} {462}
  (\bibinfo {year} {1962})}\BibitemShut {NoStop}%
\bibitem [{\citenamefont {Weeks}\ \emph {et~al.}(1997)\citenamefont {Weeks},
  \citenamefont {Vollmayr},\ and\ \citenamefont
  {Katsov}}]{weeks1997intermolecular}%
  \BibitemOpen
  \bibfield  {author} {\bibinfo {author} {\bibfnamefont {J.~D.}\ \bibnamefont
  {Weeks}}, \bibinfo {author} {\bibfnamefont {K.}~\bibnamefont {Vollmayr}}, \
  and\ \bibinfo {author} {\bibfnamefont {K.}~\bibnamefont {Katsov}},\
  }\href@noop {} {\bibfield  {journal} {\bibinfo  {journal} {Physica A}\
  }\textbf {\bibinfo {volume} {244}},\ \bibinfo {pages} {461} (\bibinfo {year}
  {1997})}\BibitemShut {NoStop}%
\bibitem [{\citenamefont {Weeks}\ \emph {et~al.}(1998)\citenamefont {Weeks},
  \citenamefont {Katsov},\ and\ \citenamefont {Vollmayr}}]{weeks1998roles}%
  \BibitemOpen
  \bibfield  {author} {\bibinfo {author} {\bibfnamefont {J.~D.}\ \bibnamefont
  {Weeks}}, \bibinfo {author} {\bibfnamefont {K.}~\bibnamefont {Katsov}}, \
  and\ \bibinfo {author} {\bibfnamefont {K.}~\bibnamefont {Vollmayr}},\
  }\href@noop {} {\bibfield  {journal} {\bibinfo  {journal} {Phys. Rev. Lett.}\
  }\textbf {\bibinfo {volume} {81}},\ \bibinfo {pages} {4400} (\bibinfo {year}
  {1998})}\BibitemShut {NoStop}%
\bibitem [{\citenamefont {Weeks}(2002)}]{weeks2002connecting}%
  \BibitemOpen
  \bibfield  {author} {\bibinfo {author} {\bibfnamefont {J.~D.}\ \bibnamefont
  {Weeks}},\ }\href@noop {} {\bibfield  {journal} {\bibinfo  {journal} {Ann.
  Rev. Phys. Chem.}\ }\textbf {\bibinfo {volume} {53}},\ \bibinfo {pages} {533}
  (\bibinfo {year} {2002})}\BibitemShut {NoStop}%
\bibitem [{Note1()}]{Note1}%
  \BibitemOpen
  \bibinfo {note} {The correlations neglected in \protect \textup {\hbox
  {\mathsurround \z@ \protect \normalfont (\ignorespaces \ref
  {eq:ACE:RPA_DFT}\unskip \@@italiccorr )}} are incorporated into the
  functional $F_{\protect \mathrm {corr}}[\rho ]$ defined in Eq.~(3.4.11) of
  Ref.\ \protect \rev@citealpnum {hansen2013theory}.}\BibitemShut {Stop}%
\bibitem [{Note2()}]{Note2}%
  \BibitemOpen
  \bibinfo {note} {There are, of course, other theories that enforce the
  hard-core condition on $g(r)$. The well-known Mean-Spherical Approximation
  (MSA)\cite {hansen2013theory, lebowitz1966mean} enforces this and sets
  $c^{(2)}(r) = -\beta v(r)$ outside the core. Perhaps more pertinent to our
  present discussion is the Optimized RPA (ORPA).\cite {hansen2013theory,
  andersen1972roles} The ORPA invokes the RPA closure \protect \textup {\hbox
  {\mathsurround \z@ \protect \normalfont (\ignorespaces \ref
  {eq:ACE:RPA_from_OZ}\unskip \@@italiccorr )}} but, in addition, seeks to vary
  the perturbation potential $v_1(r)$ inside the hard core, where this is not
  uniquely defined, so that $g(r)=0$. This constraint is equivalent to
  requiring the functional derivative of the RPA free energy with respect to
  $v_1(r)$ to be zero inside the hard core. In our present approach, there is
  no attempt to minimize the free energy with respect to the perturbation
  potential, so for $r>\sigma $, $g(r)$ depends weakly on the choice of
  $v_1(r)$ inside the core. There is no reason to expect the MF DFT to perform
  as well as the ORPA -- at least for a Lennard-Jones type fluid at high
  densities where the ORPA is known to be very accurate.\cite
  {hansen2013theory, andersen1972roles} \par For completeness, we also mention
  the Reference HNC (RHNC), see e.g.\ Ref.~\protect \rev@citealpnum
  {hansen2013theory}, which improves upon the HNC integral equation by
  approximating the bridge function by that of a reference system, usually
  hard-spheres. Although the integral equation \protect \textup {\hbox
  {\mathsurround \z@ \protect \normalfont (\ignorespaces \ref
  {eq:ACE:OZ-like}\unskip \@@italiccorr )}} that emerges from our MF DFT shares
  a feature of the RHNC, in that $B(r)$ is replaced by $B_0(r)$, this is
  implemented only in one part of the right hand side. There is no reason to
  expect the MF DFT to be as accurate as the RHNC.}\BibitemShut {Stop}%
\bibitem [{\citenamefont {Percus}(1982)}]{percus1982one}%
  \BibitemOpen
  \bibfield  {author} {\bibinfo {author} {\bibfnamefont {J.~K.}\ \bibnamefont
  {Percus}},\ }\href@noop {} {\bibfield  {journal} {\bibinfo  {journal} {J.
  Stat. Phys.}\ }\textbf {\bibinfo {volume} {28}},\ \bibinfo {pages} {67}
  (\bibinfo {year} {1982})}\BibitemShut {NoStop}%
\bibitem [{\citenamefont {Takahashi}(1942)}]{takahashi1942simple}%
  \BibitemOpen
  \bibfield  {author} {\bibinfo {author} {\bibfnamefont {H.}~\bibnamefont
  {Takahashi}},\ }\href@noop {} {\bibfield  {journal} {\bibinfo  {journal}
  {Proc. Phys.-Math. Soc. Jpn}\ }\textbf {\bibinfo {volume} {24}},\ \bibinfo
  {pages} {60} (\bibinfo {year} {1942})}\BibitemShut {NoStop}%
\bibitem [{\citenamefont {Brader}\ and\ \citenamefont
  {Evans}(2002)}]{brader2002exactly}%
  \BibitemOpen
  \bibfield  {author} {\bibinfo {author} {\bibfnamefont {J.~M.}\ \bibnamefont
  {Brader}}\ and\ \bibinfo {author} {\bibfnamefont {R.}~\bibnamefont {Evans}},\
  }\href@noop {} {\bibfield  {journal} {\bibinfo  {journal} {Physica A}\
  }\textbf {\bibinfo {volume} {306}},\ \bibinfo {pages} {287} (\bibinfo {year}
  {2002})}\BibitemShut {NoStop}%
\bibitem [{\citenamefont {Archer}\ and\ \citenamefont
  {Evans}(2013)}]{archer2013relationship}%
  \BibitemOpen
  \bibfield  {author} {\bibinfo {author} {\bibfnamefont {A.~J.}\ \bibnamefont
  {Archer}}\ and\ \bibinfo {author} {\bibfnamefont {R.}~\bibnamefont {Evans}},\
  }\href@noop {} {\bibfield  {journal} {\bibinfo  {journal} {J. Chem. Phys.}\
  }\textbf {\bibinfo {volume} {138}},\ \bibinfo {pages} {014502} (\bibinfo
  {year} {2013})}\BibitemShut {NoStop}%
\bibitem [{\citenamefont {Dijkstra}\ \emph {et~al.}(1999)\citenamefont
  {Dijkstra}, \citenamefont {Brader},\ and\ \citenamefont
  {Evans}}]{dijkstra1999phase}%
  \BibitemOpen
  \bibfield  {author} {\bibinfo {author} {\bibfnamefont {M.}~\bibnamefont
  {Dijkstra}}, \bibinfo {author} {\bibfnamefont {J.~M.}\ \bibnamefont
  {Brader}}, \ and\ \bibinfo {author} {\bibfnamefont {R.}~\bibnamefont
  {Evans}},\ }\href@noop {} {\bibfield  {journal} {\bibinfo  {journal} {J.
  Phys.: Condens. Matter}\ }\textbf {\bibinfo {volume} {11}},\ \bibinfo {pages}
  {10079} (\bibinfo {year} {1999})}\BibitemShut {NoStop}%
\bibitem [{\citenamefont {Percus}(1976)}]{percus1976equilibrium}%
  \BibitemOpen
  \bibfield  {author} {\bibinfo {author} {\bibfnamefont {J.~K.}\ \bibnamefont
  {Percus}},\ }\href@noop {} {\bibfield  {journal} {\bibinfo  {journal} {J.
  Stat. Phys.}\ }\textbf {\bibinfo {volume} {15}},\ \bibinfo {pages} {505}
  (\bibinfo {year} {1976})}\BibitemShut {NoStop}%
\bibitem [{Note3()}]{Note3}%
  \BibitemOpen
  \bibinfo {note} {\protect Note that for $\lambda =1$,
  $S_{\protect \mathrm {MF}\lambda }(k)$, introduced in Eq.~(41) of
  Ref.~\protect \rev@citealpnum {archer2013relationship}, is identical to
  $S_\protect \mathrm {RPA}(k)$ defined here.}\BibitemShut {Stop}%
\bibitem [{\citenamefont {Oettel}(2005)}]{oettel2005integral}%
  \BibitemOpen
  \bibfield  {author} {\bibinfo {author} {\bibfnamefont {M.}~\bibnamefont
  {Oettel}},\ }\href@noop {} {\bibfield  {journal} {\bibinfo  {journal} {J.
  Phys.: Condens. Matter}\ }\textbf {\bibinfo {volume} {17}},\ \bibinfo {pages}
  {429} (\bibinfo {year} {2005})}\BibitemShut {NoStop}%
\bibitem [{\citenamefont {Rodgers}\ and\ \citenamefont
  {Weeks}(2008)}]{rodgers2008local}%
  \BibitemOpen
  \bibfield  {author} {\bibinfo {author} {\bibfnamefont {J.~M.}\ \bibnamefont
  {Rodgers}}\ and\ \bibinfo {author} {\bibfnamefont {J.~D.}\ \bibnamefont
  {Weeks}},\ }\href@noop {} {\bibfield  {journal} {\bibinfo  {journal} {J.
  Phys.: Condens. Matter}\ }\textbf {\bibinfo {volume} {20}},\ \bibinfo {pages}
  {494206} (\bibinfo {year} {2008})}\BibitemShut {NoStop}%
\bibitem [{\citenamefont {Remsing}\ \emph {et~al.}(2016)\citenamefont
  {Remsing}, \citenamefont {Liu},\ and\ \citenamefont
  {Weeks}}]{remsing2016long}%
  \BibitemOpen
  \bibfield  {author} {\bibinfo {author} {\bibfnamefont {R.~C.}\ \bibnamefont
  {Remsing}}, \bibinfo {author} {\bibfnamefont {S.}~\bibnamefont {Liu}}, \ and\
  \bibinfo {author} {\bibfnamefont {J.~D.}\ \bibnamefont {Weeks}},\ }\href@noop
  {} {\bibfield  {journal} {\bibinfo  {journal} {Proc. Natl Acad. Sci. USA}\
  }\textbf {\bibinfo {volume} {113}},\ \bibinfo {pages} {2819} (\bibinfo {year}
  {2016})}\BibitemShut {NoStop}%
\bibitem [{\citenamefont {Ramakrishnan}\ and\ \citenamefont
  {Yussouff}(1979)}]{ramakrishnan1979first}%
  \BibitemOpen
  \bibfield  {author} {\bibinfo {author} {\bibfnamefont {T.~V.}\ \bibnamefont
  {Ramakrishnan}}\ and\ \bibinfo {author} {\bibfnamefont {M.}~\bibnamefont
  {Yussouff}},\ }\href@noop {} {\bibfield  {journal} {\bibinfo  {journal}
  {Phys. Rev. B}\ }\textbf {\bibinfo {volume} {19}},\ \bibinfo {pages} {2775}
  (\bibinfo {year} {1979})}\BibitemShut {NoStop}%
\bibitem [{Note4()}]{Note4}%
  \BibitemOpen
  \bibinfo {note} {Note that minimizing the grand potential corresponding to
  \protect \textup {\hbox {\mathsurround \z@ \protect \normalfont
  (\ignorespaces \ref {eq:ACE:RY_DFT}\unskip \@@italiccorr )}} within the test
  particle procedure leads to the HNC closure for the uniform fluid, provided
  the direct correlation function is determined self-consistently using the OZ
  equation\cite {evans1992density, hansen2013theory}}\BibitemShut {NoStop}%
\bibitem [{\citenamefont {Evans}\ \emph {et~al.}(1983)\citenamefont {Evans},
  \citenamefont {Tarazona},\ and\ \citenamefont {Marconi}}]{evans1983failure}%
  \BibitemOpen
  \bibfield  {author} {\bibinfo {author} {\bibfnamefont {R.}~\bibnamefont
  {Evans}}, \bibinfo {author} {\bibfnamefont {P.}~\bibnamefont {Tarazona}}, \
  and\ \bibinfo {author} {\bibfnamefont {U.~M.~B.}\ \bibnamefont {Marconi}},\
  }\href@noop {} {\bibfield  {journal} {\bibinfo  {journal} {Mol. Phys.}\
  }\textbf {\bibinfo {volume} {50}},\ \bibinfo {pages} {993} (\bibinfo {year}
  {1983})}\BibitemShut {NoStop}%
\bibitem [{\citenamefont {Evans}\ and\ \citenamefont
  {Marconi}(1986)}]{evans1986comment}%
  \BibitemOpen
  \bibfield  {author} {\bibinfo {author} {\bibfnamefont {R.}~\bibnamefont
  {Evans}}\ and\ \bibinfo {author} {\bibfnamefont {U.~M.~B.}\ \bibnamefont
  {Marconi}},\ }\href@noop {} {\bibfield  {journal} {\bibinfo  {journal} {Phys.
  Rev. A}\ }\textbf {\bibinfo {volume} {34}},\ \bibinfo {pages} {3504}
  (\bibinfo {year} {1986})}\BibitemShut {NoStop}%
\bibitem [{\citenamefont {Lebowitz}\ and\ \citenamefont
  {Percus}(1966)}]{lebowitz1966mean}%
  \BibitemOpen
  \bibfield  {author} {\bibinfo {author} {\bibfnamefont {J.~L.}\ \bibnamefont
  {Lebowitz}}\ and\ \bibinfo {author} {\bibfnamefont {J.~K.}\ \bibnamefont
  {Percus}},\ }\href@noop {} {\bibfield  {journal} {\bibinfo  {journal} {Phys.
  Rev.}\ }\textbf {\bibinfo {volume} {144}},\ \bibinfo {pages} {251} (\bibinfo
  {year} {1966})}\BibitemShut {NoStop}%
\bibitem [{\citenamefont {Andersen}\ \emph {et~al.}(1972)\citenamefont
  {Andersen}, \citenamefont {Chandler},\ and\ \citenamefont
  {Weeks}}]{andersen1972roles}%
  \BibitemOpen
  \bibfield  {author} {\bibinfo {author} {\bibfnamefont {H.~C.}\ \bibnamefont
  {Andersen}}, \bibinfo {author} {\bibfnamefont {D.}~\bibnamefont {Chandler}},
  \ and\ \bibinfo {author} {\bibfnamefont {J.~D.}\ \bibnamefont {Weeks}},\
  }\href@noop {} {\bibfield  {journal} {\bibinfo  {journal} {J. Chem. Phys.}\
  }\textbf {\bibinfo {volume} {56}},\ \bibinfo {pages} {3812} (\bibinfo {year}
  {1972})}\BibitemShut {NoStop}%
\end{thebibliography}

%

\end{document}